%% file: ppbar.tex
\documentclass[12pt,a4paper,dvips]{article}
\usepackage{a4p}
\usepackage{cite,mcite} 
\usepackage{graphicx}
\usepackage{physics}
\usepackage{l3_title,ifthen,Lep}
\usepackage{epsfig,float}
\journalname{Phys. Lett. B}
\preprint{2003-014}
\date{March 26, 2003}

\def\ra{\rightarrow }

\def\epem{\mbox{e}^+\mbox{e}^- }

\def\ra{\rightarrow } 
\def\epem{\mbox{e}^+\mbox{e}^- } 
\def\la{\Lambda } 
\def\ppbar{\rm p \overline{p}}

\def\spt{(\sum \vec{p_t})^2} 
\def\wgg{W_{\gamma \gamma}}
\newlength{\capindent}
\newlength{\capwidth}
\newlength{\figwidth}
\newcommand{\icaption}[2][!*!,!]{\hspace*{\capindent}%
  \begin{minipage}{\capwidth}
    \ifthenelse{\equal{#1}{!*!,!}}%
      {\caption{#2}}%
      {\caption[#1]{#2}}
  \end{minipage}}
\begin{document}
\begin{titlepage}

\title{ 
Proton-Antiproton Pair Production in \\ Two-Photon Collisions at LEP}

\author{The L3 Collaboration}

%
%
%
\begin{abstract}

The reaction $\epem \ra \epem \ppbar$ is studied with the L3 detector
at LEP. The analysis is based on data collected at $\epem$
center-of-mass energies from 183 GeV to 209 GeV, corresponding to an
integrated luminosity of 667 pb$^{-1}$. The $\gamma \gamma \ra \ppbar$
differential cross section is measured in the range of the two-photon
center-of-mass energy from 2.1 GeV to 4.5 GeV. The results are
compared to the predictions of the three-quark and quark-diquark
models.

\end{abstract}

\submitted

\end{titlepage}

\section{Introduction}

Electron-positron colliders are a suitable place for the study of
two-photon interactions via the process $\epem \ra \epem \gamma^*
\gamma^* \ra \epem X$, where $\gamma^*$ denotes a virtual photon. The
outgoing electron and positron carry almost the full beam energy and
are usually undetected, due to their small transverse momenta. The
final state $X$ has, therefore, a low mass as compared to the $\epem$
center-of-mass energy, $ \sqrt{s}$, and its transverse momentum is
almost zero. The small photon virtuality allows the extraction of the
cross section \mbox{$\sigma(\gamma \gamma \ra X)$} for real photon
collisions, once the photon flux is calculated by QED \cite{th1}.

Calculations of the cross section \mbox{$\gamma \gamma \ra $ \sl
baryon antibaryon} were performed using the hard scattering approach
of Brodsky and Lepage \cite{th2}. In this formalism, the process is
factorized into a perturbative $\gamma \gamma \ra \rm q \overline{q}$
amplitude and a non-perturbative part described by the quark
distribution functions of the baryon. Such calculations with
three-quark distribution functions \cite{th3,th4} yielded results
about one order of magnitude below early $\gamma \gamma \ra \ppbar$
measurements \cite{prev1} for $\gamma \gamma$ center-of-mass energies,
$W_{\gamma \gamma}$, less than 3 GeV. The quark-diquark model
\cite{th5} was proposed as a possible way to model non-perturbative
effects. Within this model, the partonic structure of the baryon is
described by a quark-diquark system rather than three quarks. The
composite nature of diquarks is taken into account by form factors
ensuring that the three-quark model is recovered at an asymptotically
large momentum transfer.

This Letter presents a study of the differential cross section of the
reaction $\gamma \gamma \ra \rm p \overline{p}$ in the $W_{\gamma
\gamma}$ range from 2.1 GeV to 4.5 GeV. The data sample corresponds to
a total integrated luminosity of 667 pb$^{-1}$ collected with the L3
detector \cite{det} at \mbox{$\sqrt{s}= 183 - 209$ GeV}. The analysis
is based on the central tracking system and the high resolution BGO
electromagnetic calorimeter. The events are selected by the track
triggers \cite{trig1}.

The $\rm p \overline{p}$ final state in two-photon collisions was
previously studied at lower $\sqrt{s}$ \cite{prev1,prev2} and recently
at LEP \cite{opal} with lower statistics. The present study extends
the $\W_{\gamma \gamma}$ range. Our results are compared to these
experiments and to recent theoretical predictions of the quark-diquark
model \cite{th6}.

\section{Monte Carlo simulation}

Monte Carlo events for the \mbox{$\gamma \gamma \ra \rm p
\overline{p}$} reaction, as well as for the background processes
\mbox{$\gamma \gamma \ra \rm K^+ K^-$} and \mbox{$\gamma \gamma \ra
\rm \pi^+ \pi^-$}, are generated with the PC Monte Carlo program
\cite{gen1}, for each beam energy, within the formalism of Reference
\citen{th1}. The \mbox{$\rm e^+e^- \rightarrow \rm
e^+e^-\tau^+\tau^-$}, \mbox{$\rm e^+e^- \rightarrow \rm e^+e^-
\mu^+\mu^-$} and \mbox{$\rm e^+e^- \rightarrow \rm e^+e^-e^+e^-$}
background processes are simulated with the DIAG36 \cite{gen4}
generator which includes the full set of ${\cal O}(\alpha^4)$ QED
diagrams. The generated events are passed through the full L3 detector
simulation using the GEANT \cite{gen2} and GEISHA \cite{gen3} programs
and are reconstructed with the same programs as the data. Time
dependent detector inefficiencies, as monitored during the data taking
period, are taken into account.

\section{Proton-antiproton event selection}

In order to match the experimental acceptance with the range of the
theoretical predictions, the data are analysed only in the region $|
\cos \theta^* | < 0.6$, where $\theta^*$ is the center-of-mass
production angle of the proton. Events are first selected by requiring
two well-reconstructed tracks of opposite charge. The track selection
criteria are:

\begin{itemize}

\item a distance of closest approach to the interaction point less
than 3 mm in the plane transverse to the beam direction,

\item at least 30 hits, out of a maximum of 62, in the tracking
chamber,

\item a matched energy cluster in the electromagnetic calorimeter,

\item a transverse momentum, $p_t$, greater than 400 MeV, to ensure a
high trigger efficiency and electron rejection.

\end{itemize}

The identification of $\gamma \gamma \ra \rm p \overline{p}$ events is
mainly based on three artificial neural networks, used to separate
antiprotons from $\rm e^-$, $\rm \mu^-$ and $\rm h^-$, where $\rm h^-$
represents either a $\pi^-$ or $\rm K^-$. Each neural network consists
of five input nodes, a single layer of five hidden neurons and one
output neuron. The following measured quantities are associated with
the five input nodes:

\begin{itemize}

\item the momentum of the antiproton,

\item the probability for the proton mass hypothesis based on the mean
energy loss $dE/dx$ measured in the tracker,

\item the ratio $ E_t/p_t$, where $ E_t$ is the transverse energy
measured in the electromagnetic calorimeter,

\item the number of BGO crystals in the calorimetric cluster
      associated with the antiproton,

\item the ratio between the energy deposited in the central crystal of
      the cluster and the sum of the energies deposited in the
      three-by-three matrix of crystals around it.

\end{itemize}

The two last variables exploit the typical signature of an antiproton
annihilating in the BGO electromagnetic calorimeter: a broad shower
spanning several crystals, as opposed to narrow showers for electrons,
minimum ionisation deposition for muons and compact showers for low
energy hadrons. Each neural network is trained with the corresponding
sample from Monte Carlo simulations so that its output value is close
to one for antiprotons.

\begin{figure}
\begin{center}

\includegraphics[height=9cm]{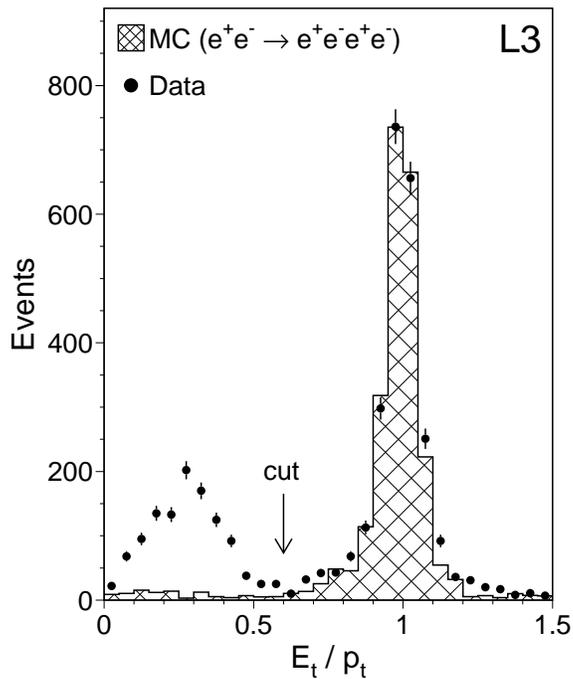}
 \end{center}
\caption {The ratio between the transverse energy deposited in the electromagnetic calorimeter, $E_t$, and the transverse momentum, $p_t$, for the proton candidate after the antiproton selection. The $\epem \ra \epem \epem$ Monte Carlo prediction is superimposed on the data.}
\label{etptg}
\end{figure}

A particle is identified as an antiproton after a cut on the output of
the three neural networks. The efficiency for antiproton detection is
found to be 74\%. Even though the $\rm e^-$ neural network rejects
more than 98\% of the electrons, the $\epem \ra \epem \epem$ cross
section is more than two orders of magnitude greater than the $\epem
\ra \epem \ppbar$ cross section and the electron contamination still
remains important. To reduce this background, the ratio $E_t/p_t$ of
the proton candidate is required to be less than 0.6, as shown in
Figure \ref{etptg}. In addition, the confidence level that the $dE/dx$
measurement is consistent with a proton must be more than 5\%. These
cuts eliminate more than 95\% of the remaining $\rm e^+e^- \ra
e^+e^-e^+e^-$ background.

\section{Exclusive ${\bf \gamma \gamma} \rightarrow \mathbf{p \overline{p}} $ event selection}

To select exclusive $\gamma \gamma \ra \rm p \overline{p}$ events the following
further cuts are applied:

\begin{itemize}

\item Events with a photon candidate outside a cone of 36 degrees
half-opening angle with respect to the antiproton direction are
rejected. A photon candidate is defined as a shower in the
electromagnetic calorimeter with at least two adjacent crystals, an
energy greater than 100 MeV and no charged tracks within a cone of 200
mrad half-opening angle.

\item The square of the total transverse momentum of the
      proton-antiproton pair $(\sum \vec{p_t})^2$ must be less than
      0.1 GeV$^2$. This cut reduces inclusive background events of the
      type ${\epem \ra \epem \rm p \overline{p}} X$, where $X$
      represents one or more unobserved particles.

\end{itemize}

\begin{figure}
\begin{center}
\includegraphics[height=9cm]{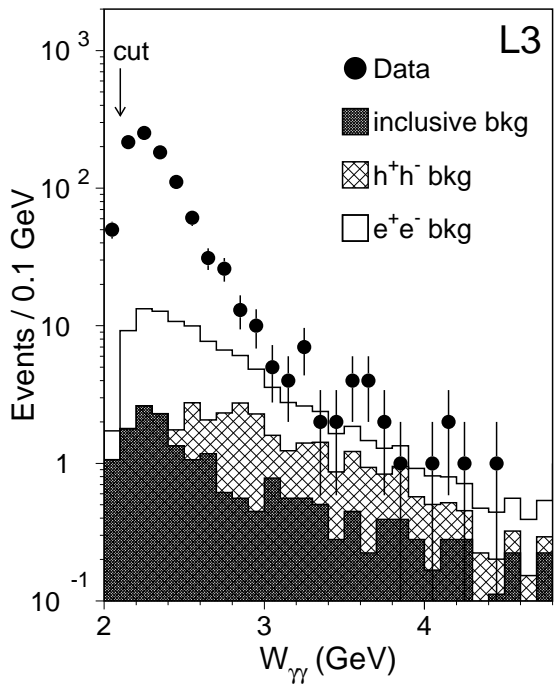}
 \end{center}
\caption {The effective mass of the $\rm p \overline{p}$ pair, $\wgg$, for the 989 selected events. The 938 events at the right of the cut are used in the subsequent analysis.}
\label{mass}
\end{figure}

These cuts yield a total number of 989 selected events. Figure
\ref{mass} shows the distribution of the effective mass of the
$\ppbar$ pair, identified with $\wgg$. It is obtained by assigning the
proton mass to the two tracks.

The exclusive background is estimated by processing the corresponding
Monte Carlo events with the same analysis cuts. The \mbox{$ \gamma
\gamma \ra \pi^+ \pi^-$} and \mbox{$\rm \gamma \gamma \ra K^+ K^-$}
cross sections given in Reference \citen{xsct} are used. This
background is found to be negligible in the region $\wgg < 2.5$ GeV
and increases up to 25\% above 2.5 GeV. The $\epem \ra \epem \epem $
background varies from 5\% in the low mass region to 30\% in the high
mass region. The $\epem \ra \epem \mu^+ \mu^-$ contamination is less
than 0.2\% and neglected.  The contamination of inclusive channels and
of $\epem \ra \epem \tau^+ \tau^-$ is estimated by fitting the tail of
the $\spt$ distribution. The extrapolation to the region $\spt < 0.1$
GeV$^2$ gives an average background level of \mbox{(2 $\pm$ 1)\%}. The
sum of all backgrounds is subtracted bin by bin for the cross section
determination. The background composition of each bin is detailed in
Table \ref{bkgtable}.

\section{Cross section measurements}

\begin{figure}
\begin{center}
\includegraphics[height=10cm]{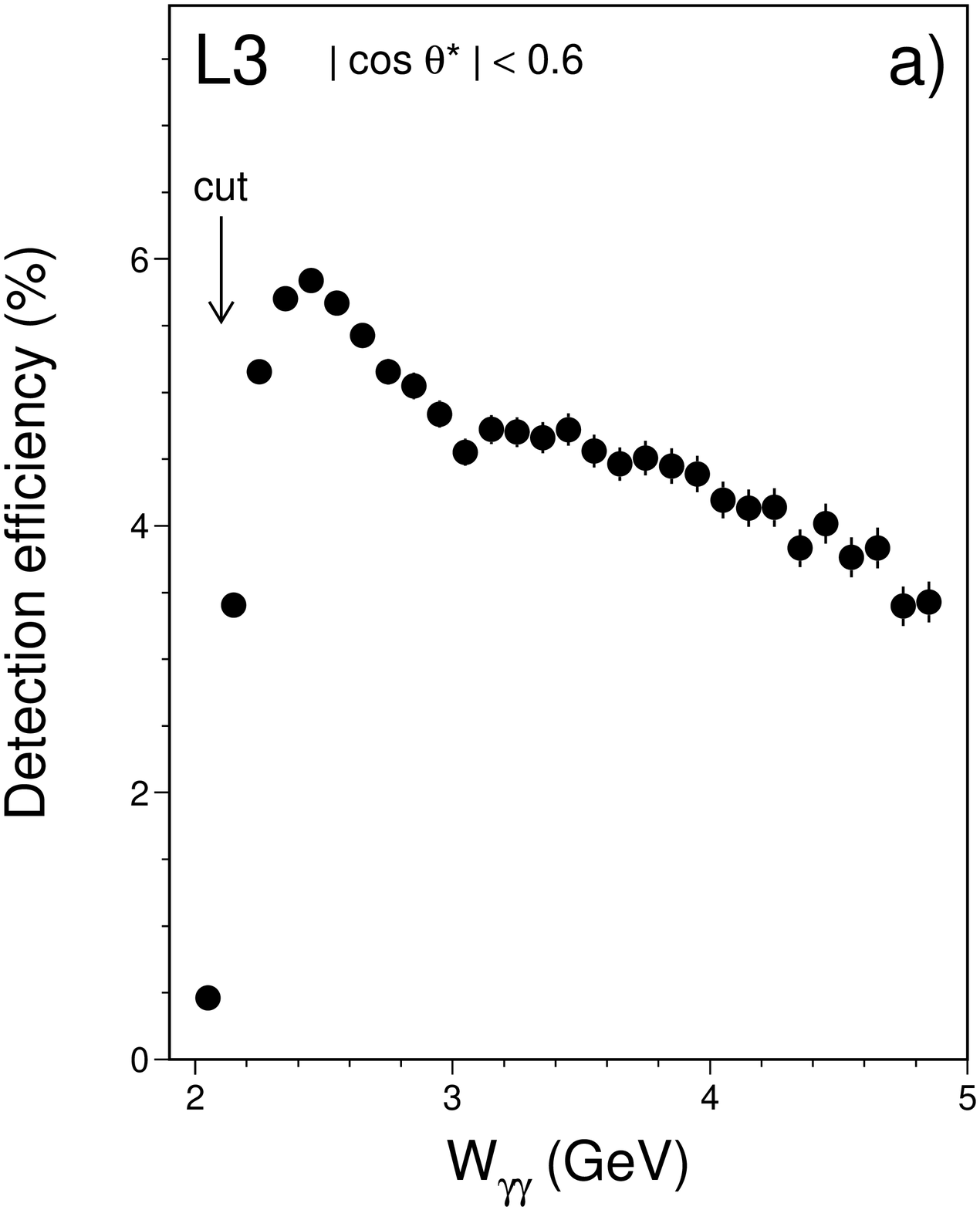}
\includegraphics[height=10cm]{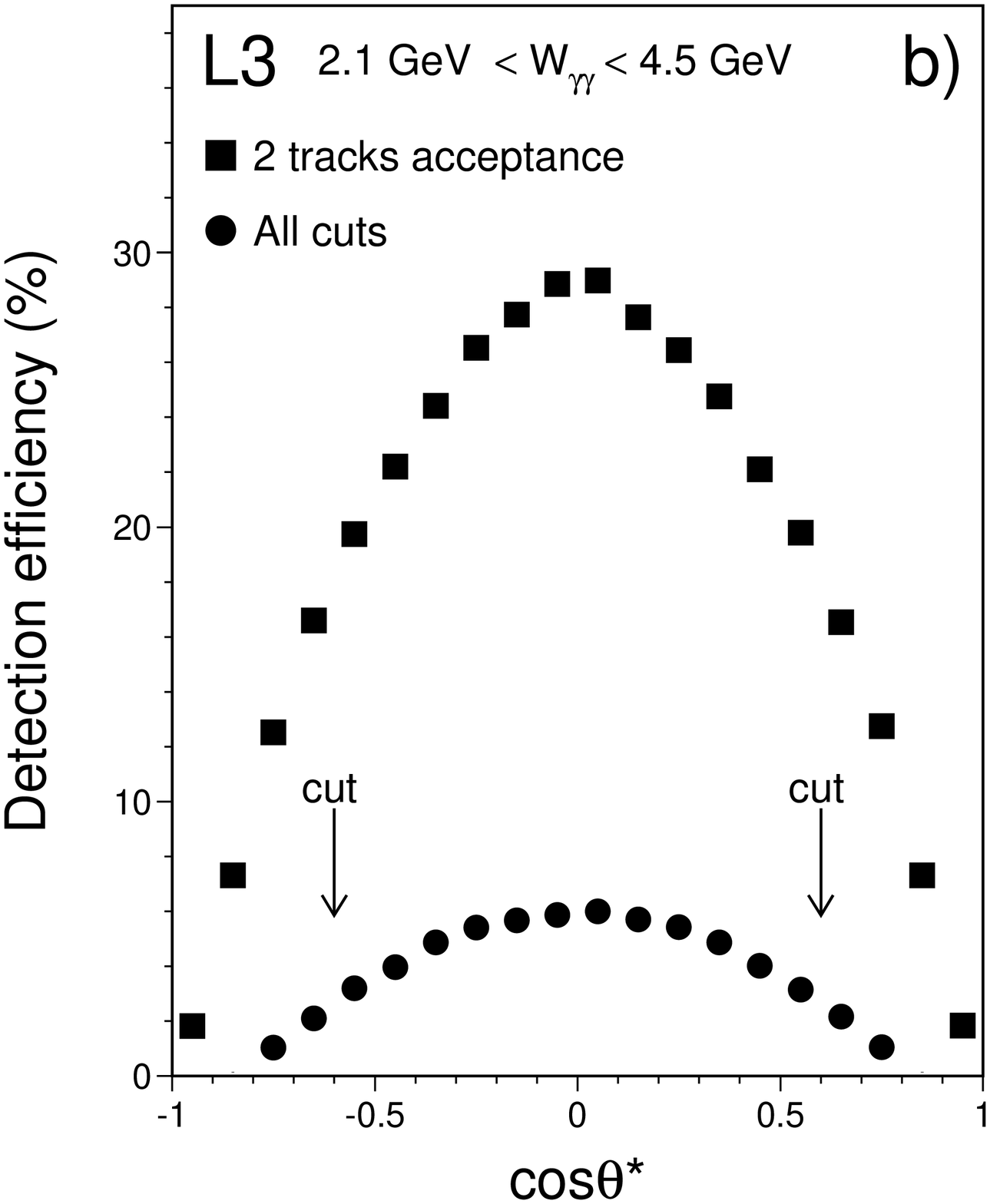}
 
\end{center}
\caption {The selection efficiency as a function of a) $\wgg$ for $| \cos \theta^* | < 0.6$ and b) $\cos \theta^*$ (full circles) for 2.1 GeV $ < \wgg <$ 4.5 GeV. The full squares indicate the efficiency for detecting two charged tracks of opposite charge in the detector, with no further requirements.}
\label{efficiency}
\end{figure}

The detection efficiency is determined by Monte Carlo simulation in
bins of $\wgg$ and $|\cos \theta^*|$. It takes into account the track
acceptance and selection criteria, the exclusive proton-antiproton
identification criteria and the track trigger efficiency. This trigger efficiency is 91\% at 
 $\wgg =2.1 $ GeV and rises to  98\% for  $\wgg >3 $ GeV.
The higher
level trigger efficiencies are estimated from the data themselves,
using prescaled events, and range from 88\% at $\wgg =2.1 $ GeV up to 99\% for  $\wgg >3 $ GeV.
The total selection efficiency is found to be
maximum, about 6\%, at $\wgg \simeq 2.5 $ GeV and $|\cos \theta^*|
<0.1$. It decreases both at larger $|\cos \theta^*|$ and $\wgg$, due
respectively to limited angular acceptance and antiproton
identification efficiency. The average efficiencies as a function of
$\wgg$ for $|\cos \theta^*| < 0.6$ and as a function of $|\cos
\theta^*| $ for $\wgg < 4.5 $ GeV are presented in Figure
\ref{efficiency}.
   
The cross sections are evaluated in bins of $\wgg$ and
$|\cos\theta^*|$ for different values of $\sqrt{s}$. Due to the
limited selection efficiency near threshold (Figure \ref{efficiency}),
the measurement is restricted to the range $ \wgg> 2.1$ GeV. A total
of 938 events are selected after this cut. The differential cross
sections are integrated to obtain the production cross sections $\epem
\ra \epem \ppbar$ for $| \cos \theta^* | < 0.6$ and 2.1 GeV $ < \wgg<$
4.5 GeV as a function of $\sqrt{s}$. Since the results show no
significant $\sqrt{s}$ dependence they are combined into a single
measurement at $\langle \sqrt{s} \rangle =197$ GeV:

\begin{center}
$ \sigma(\epem \ra \epem \ppbar) = 26.7 \pm 0.9 \pm 2.7 \;\; \rm pb$
\end{center}

\noindent where the first uncertainty is statistical, the second systematic.

The following systematic uncertainties are considered. The uncertainty
due to the selection procedure is evaluated by varying the selection
cuts. The uncertainty on the neural network outputs is estimated by
varying the input variables according to their resolution. An
uncertainty of 50\% on the $\gamma \gamma \ra \pi^+\pi^-$ and $\gamma
\gamma \ra \rm K^+K^-$ cross sections is propagated in the background
subtraction. The uncertainty on the $\rm e^+e^-\rightarrow
e^+e^-e^+e^-$ background is estimated to be 30\%. The uncertainty of
50\% on the level of inclusive background is also included.
For low values of $\wgg$ the dominant systematic
uncertainty is about 7\% and is due to selection cut variation. It remains
the most important systematic uncertainty for $\wgg<3.2$ GeV, as large
as 30\%. For  $\wgg>3.2$ GeV, the background subtraction uncertainty
becomes the most important, rising to 55\% in the last  $\wgg$  bin.
The systematic
uncertainties shown in Table \ref{res1} are the quadratic sum of the
different sources.

The differential cross section $d\sigma(\gamma \gamma \ra \ppbar )/
d|\cos\theta^*|$ for real photon-photon collisions is extracted as a
function of $W_{\gamma \gamma}$ by dividing out the two-photon
luminosity function and extrapolating to $Q^2=0$ with a GVDM form
factor ~\cite{b14}. The luminosity functions are evaluated for each
$\sqrt{s}$ and $W_{\gamma \gamma}$ interval. The measured differential
cross sections are integrated to obtain the cross sections
$\sigma(\gamma \gamma \ra \ppbar )$. An additional uncertainty of 5\%,
due to the choice of the photon form factor, is included in the
systematics. The results are reported in Table~\ref{res1}. The present
measurement is of higher statistical precision and extends towards
higher values of $W_{\gamma \gamma}$ than the previous results
\cite{prev1,prev2,opal}. Agreement is observed within the quoted
uncertainties, except for the measurements of Reference \citen{opal}
which lie below our data in the low mass region.

\section{Discussion of the results}

\begin{figure}
\begin{center}
\includegraphics[height=12cm]{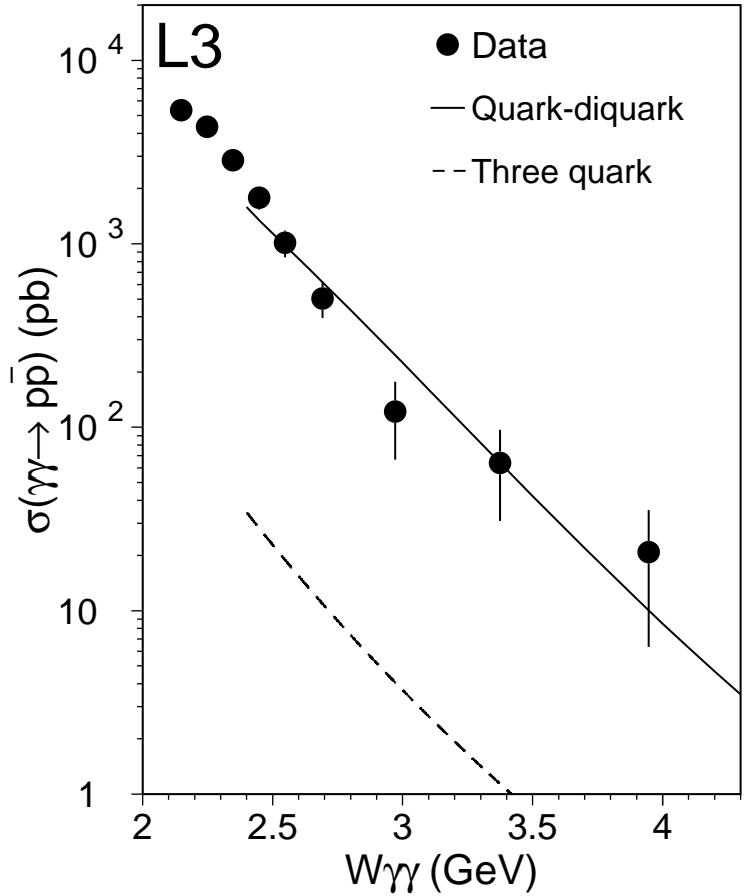} 
\end{center}
\caption {
The $\gamma \gamma \rightarrow \ppbar$ cross sections as a function of $W_{\gamma\gamma}$ for $|\cos \theta^*| < 0.6$ compared to the three-quark model calculation {\protect \cite{th3}} and to the recent quark-diquark model prediction {\protect \cite{th6}}, available for $W_{\gamma \gamma} >2.5$ GeV. Statistical and systematic uncertainties are added in quadrature.}
\label{comp}
\end{figure}

The results are compared to the predictions of the three-quark
\cite{th3} and the recent quark-diquark model \cite{th6} in
Figure~\ref{comp}. The three-quark prediction is based on the leading
order QCD calculations, using the distribution function of Chernyak
and Zhitnisky \cite{wavefct}. The quark-diquark calculation is
performed with the standard proton distribution amplitudes \cite{th6}
and includes first order corrections due to the non-vanishing proton
mass. While the shapes of the theoretical curves are quite similar,
the normalisations are significantly different. The predictions of the
three-quark model are about an order of magnitude below the
measurement, whereas the quark-diquark predictions describe the data
much better. The apparent change in the logarithmic slope of the cross
section observed in data at $\wgg$ around 3 GeV is, however, not
reproduced by this model.

\begin{figure}
\begin{center}
\includegraphics[height=8cm,angle=90]{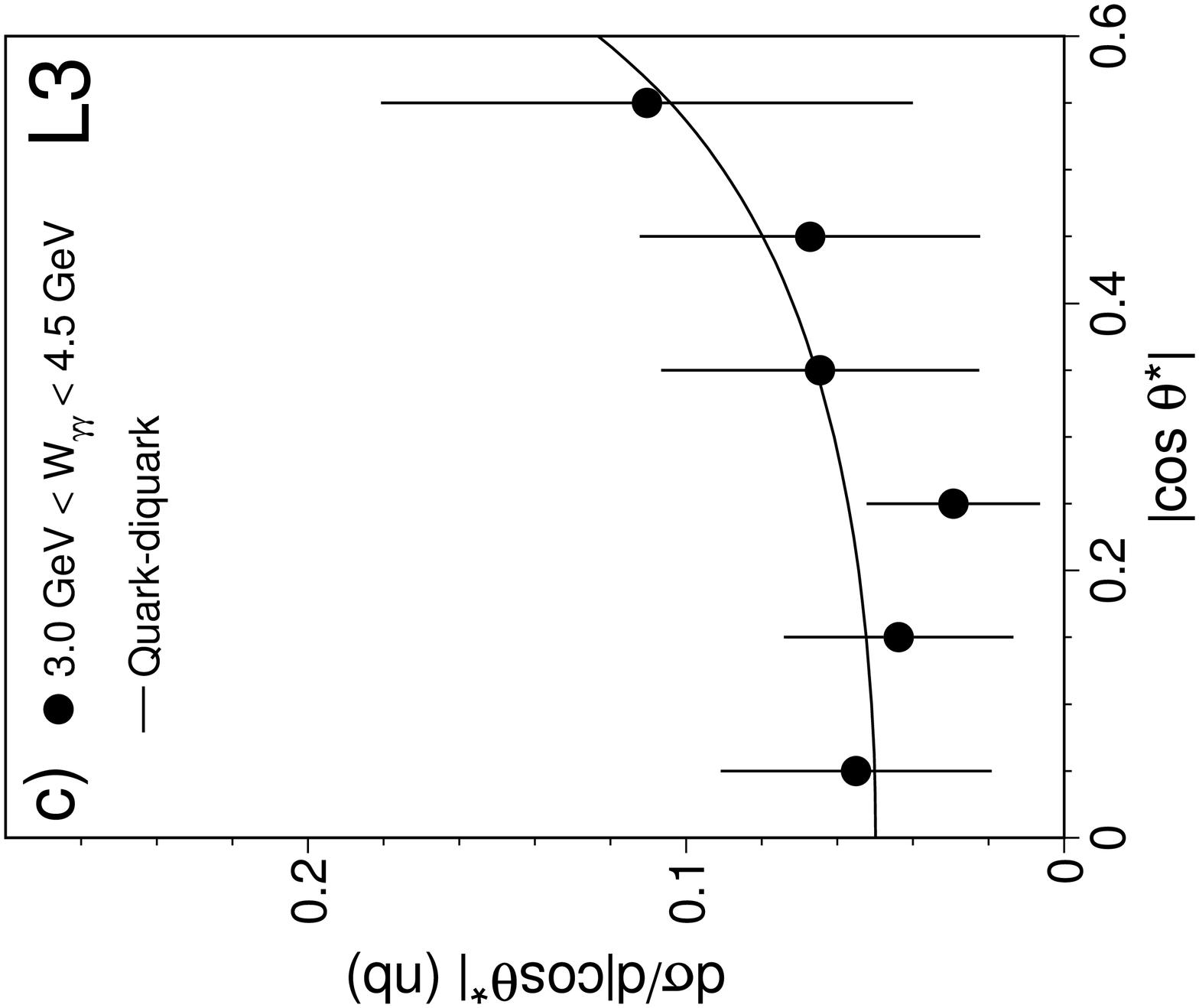} \\
\includegraphics[height=8cm,angle=90]{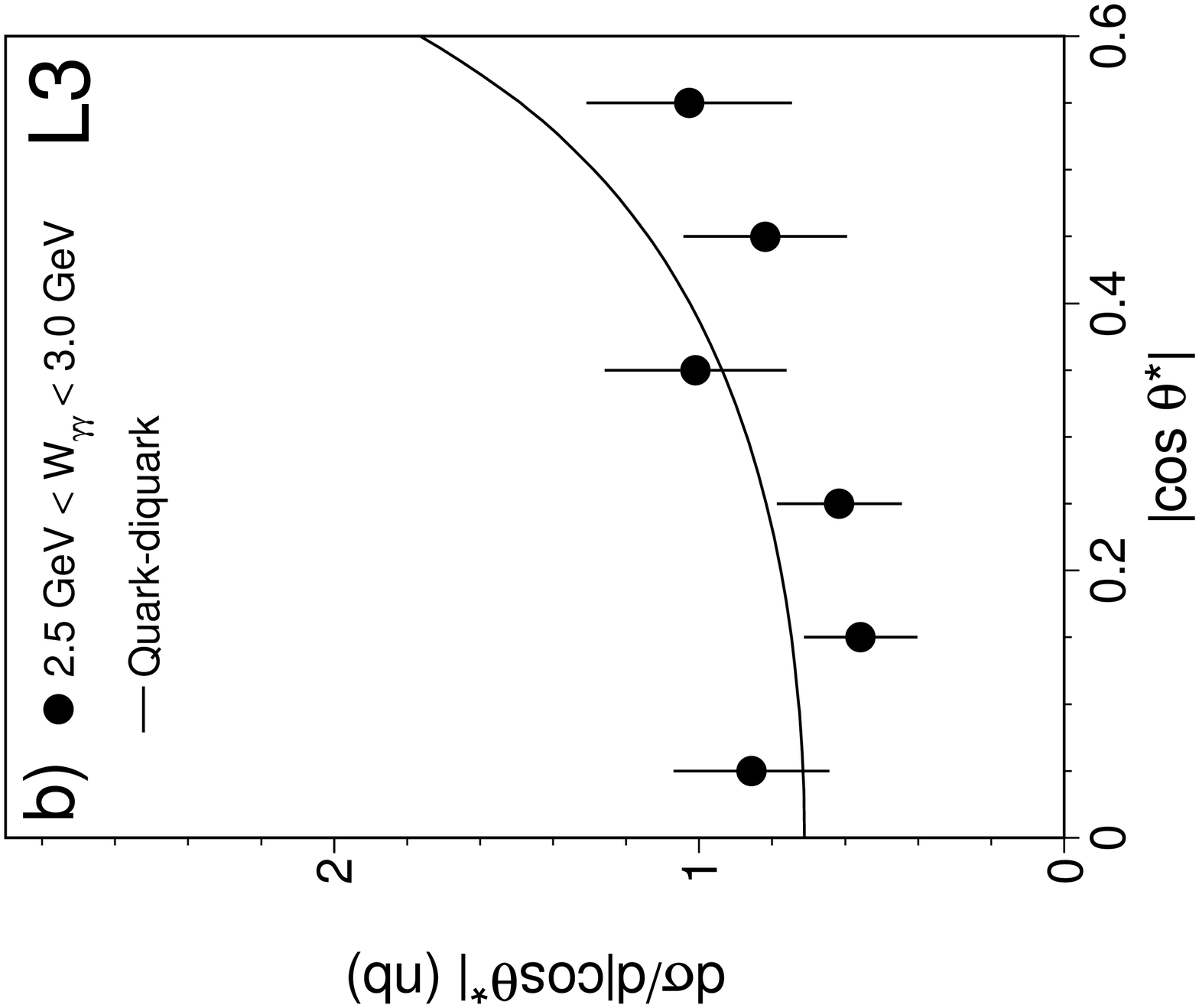} \\
\includegraphics[height=8cm,angle=90]{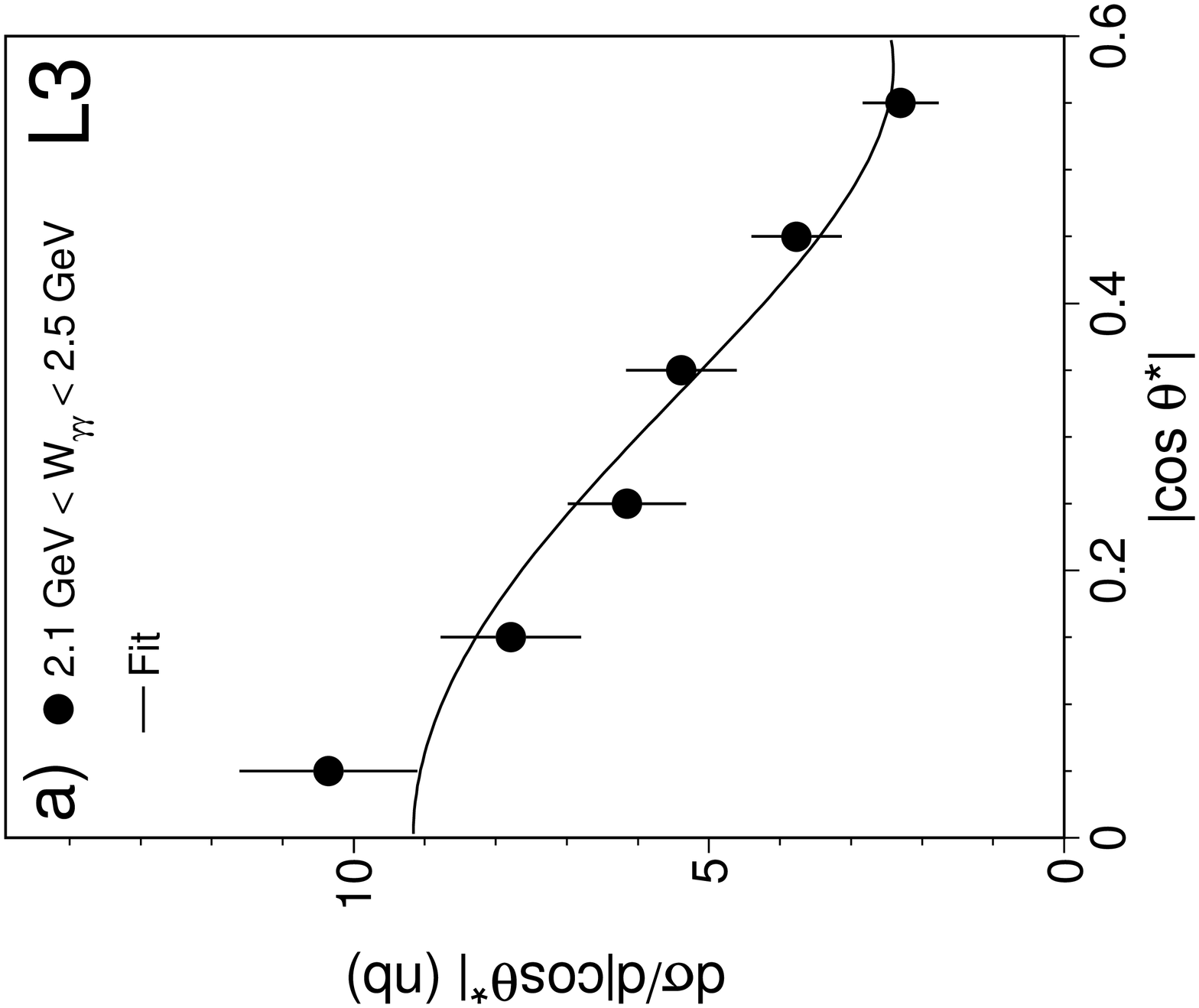}
\end{center}
\caption {The differential cross section as a function of $|\cos\theta^*|$ for a) $2.1 < W_{\gamma \gamma} < 2.5$ GeV, b) $2.5< W_{\gamma \gamma}< 3.0 $ GeV and c) $3.0 < W_{\gamma \gamma} < 4.5 $ GeV. The data are compared to the predictions of the quark-diquark model. The fit in a) is described in the text. }
\label{diffplot}
\end{figure}

In order to investigate this further, the differential cross sections
are summed separately in three mass intervals: \mbox{$2.1$ GeV $<
W_{\gamma \gamma}<2.5$ GeV}, \mbox{$2.5$ GeV $< W_{\gamma \gamma}<3.0$
GeV} and $3.0$ GeV $ < W_{\gamma \gamma} < 4.5$ GeV. The results are
reported in Table \ref{difftable} and plotted in Figure
\ref{diffplot}. A distinctive difference between the three
distributions can be observed. No prediction is available for the
diquark model for $\wgg < 2.5$ GeV, but it can be seen that the data
has a qualitatively different behaviour to the diquark predictions, as
it is strongly peaked at large angles. Assuming the presence of a
single angular momentum state in the $s$-channel of  $\gamma \gamma \ra \ppbar$
production, a fit to a single spherical
harmonic: $Y_0^0$, $Y_2^0$, $Y_2^1$ or $Y_2^2$ is performed. This does
not give a satisfactory result. An acceptable fit is obtained using
all of them, but only the $Y_2^0$ contribution is significantly
different from zero. A satisfactory fit is also obtained using only
the $Y_0^0$ and $Y_2^0$ harmonics with the fractions 8\% and 92\%
respectively, as shown in Figure \ref{diffplot}a. The intermediate
region exhibits a rather flat dependence, which partially agrees with
the model predictions. The forward peaking behaviour of the
differential cross section in the high mass interval is well
reproduced by the quark-diquark model. Only in this region, then, the
data can be described by the Brodsky-Lepage hard scattering approach.

\begin{figure}
\begin{center}
\includegraphics[height=8cm,angle=90]{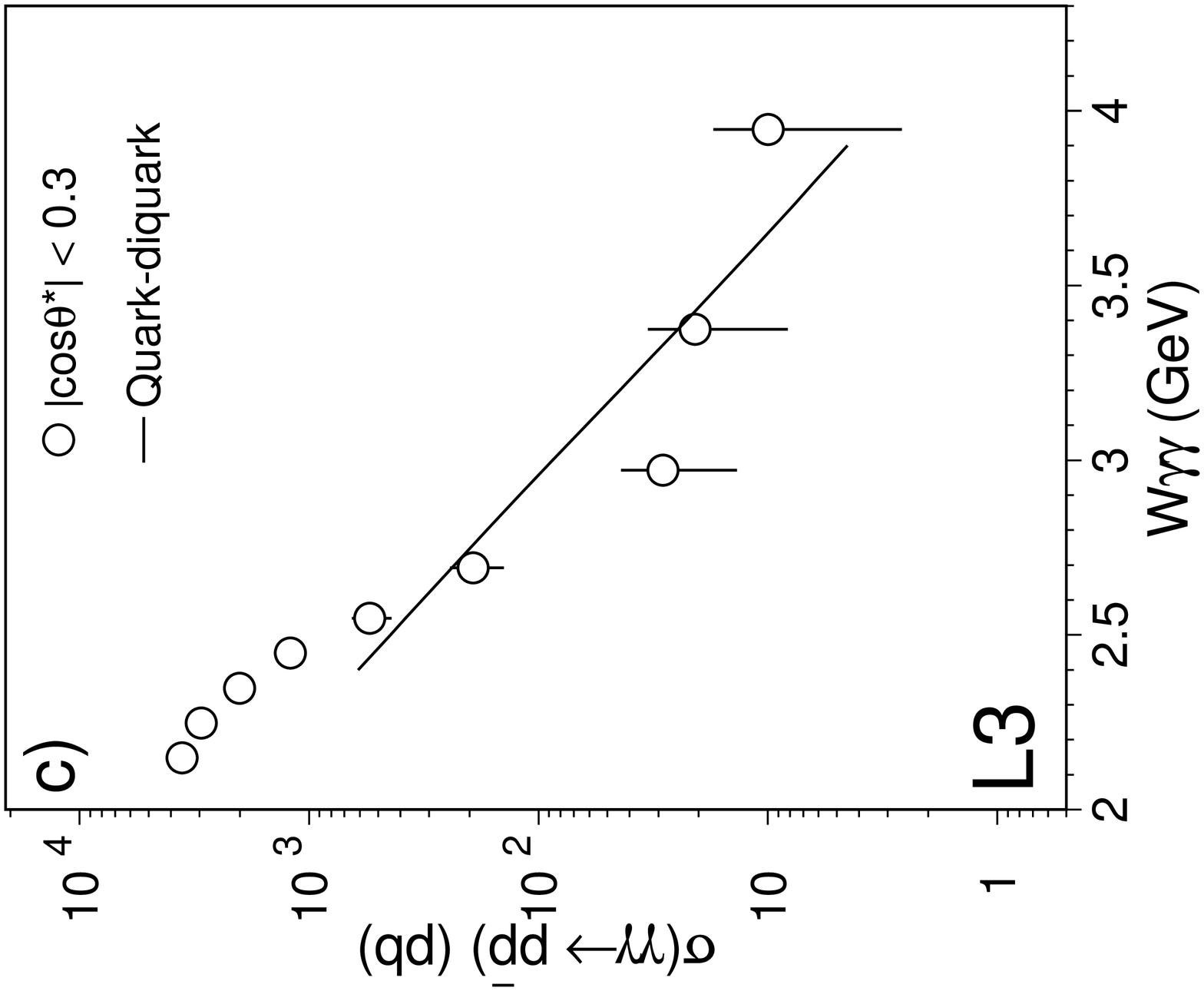} \\  
\includegraphics[height=8cm,angle=90]{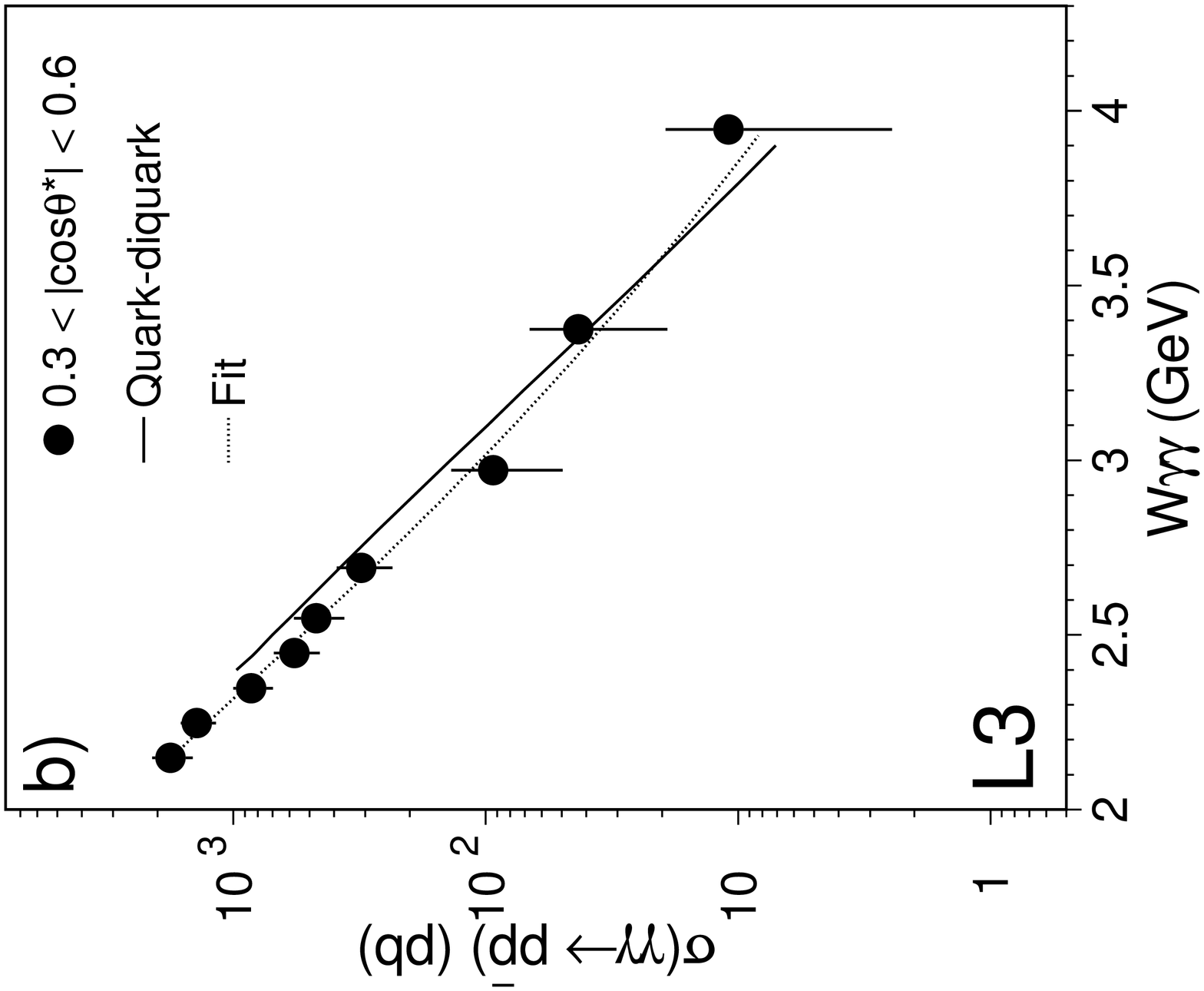} \\ 
\includegraphics[height=8cm,angle=90]{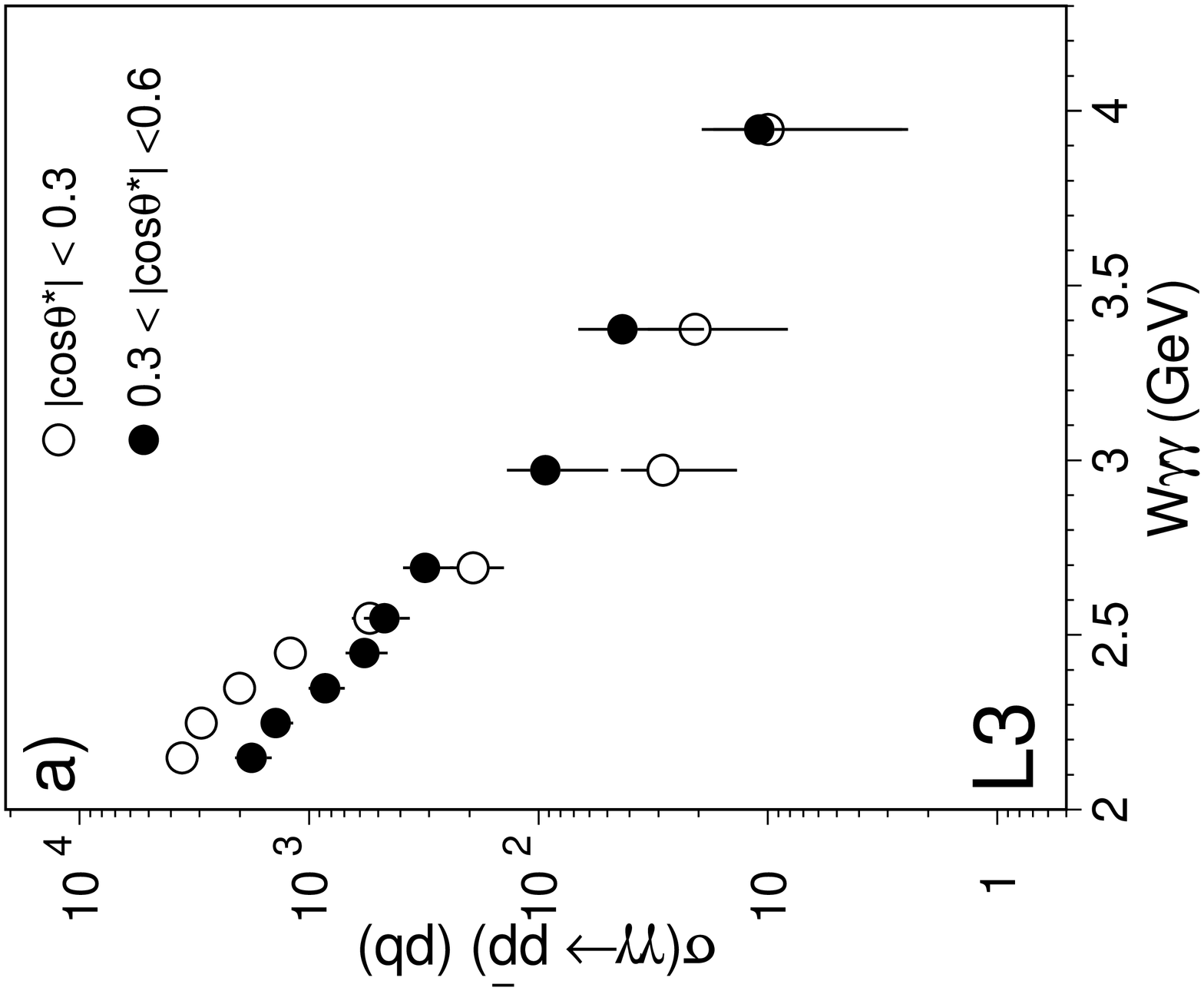} 
\end{center}
\caption {a) The $\gamma \gamma \rightarrow \ppbar$ cross section as a function of $W_{\gamma\gamma}$ for the large angle region, $|\cos \theta^*|<0.3$ (full circle), and the small angle region, $0.3<|\cos \theta^*|<0.6$ (open circle).  b) The small angle and c) the large angle cross section with the quark-diquark model predictions and the fit described in text.}
\label{sumcos}
\end{figure}

The presence of two distinct production mechanisms can also be seen in
Figure \ref{sumcos}a and Table \ref{diffwggtable} by considering
separately the cross sections as a function of $W_{\gamma \gamma}$ in
a large angle region, $|\cos \theta^*|<0.3$, and in a region
$0.3<|\cos \theta^*|<0.6$. The shape of the efficiency as a function
of $W_{\gamma \gamma}$ is similar for the two regions. The large angle
cross section dominates in the low mass region and shows a steeper
fall-off with $W_{\gamma \gamma}$ than the small angle cross
section. A fit of the form $\sigma(W) \propto W^{-n}$ to the small
angle region shown in Figure \ref{sumcos}b gives a reasonably good
description with $n = 9.8 \pm 0.3$. The data are also in agreement
with the quark-diquark model prediction in this region. On the other
hand, Figure \ref{sumcos}c, the large angle cross section cannot be
described by a simple power law behaviour for $\wgg$ less than 3 GeV
and does not follow the quark-diquark prediction. In this case, the
change of shape at 3 GeV is similar to, but more pronounced than that
in Figure \ref{comp}.

In conclusion, an accurate study of the $\gamma \gamma \ra \ppbar$
process is performed. The present results are of higher statistical
precision and extend towards higher values of $W_{\gamma \gamma}$ than
those of previous experiments. Current models are only moderately
successful in describing the observed data. These data can provide
useful inputs for predictions of other \mbox{$\gamma \gamma \ra $ \sl
baryon antibaryon} channels \cite{handbag}.

\newpage
\bibliographystyle{l3style}

%
%

\newpage
\input namelist266.tex
\newpage

%
%

\newpage

\begin{table}[H]
\begin{center}
\begin{tabular}{|c|c|c|c|} \hline 
$ \wgg$ & inclusive & $\epem$  & $\rm h^+h^-$  \\
(GeV) & (\%) & (\%) & (\%) \\\hline

2.1 $-$ 2.2 & \phantom{0}1 & \phantom{0}3 & \phantom{0}$<$1\phantom{0}\\

2.2 $-$ 2.3 & \phantom{0}1 & \phantom{0}4 & \phantom{0}$<$1\phantom{0}\\

2.3 $-$ 2.4 & \phantom{0}1 & \phantom{0}6  & \phantom{0}$<$1\phantom{0}\\

2.4 $-$ 2.5 & \phantom{0}1 & \phantom{0}8  & \phantom{0}$<$1\phantom{0}\\

2.5 $-$ 2.6 & \phantom{0}2 & 12            & \phantom{0$<$}3\phantom{0}\\
 
2.6 $-$ 2.8 & \phantom{0}3 & 18            & \phantom{0$<$}5\phantom{0} \\
 
2.8 $-$ 3.2 & \phantom{0}7 & 29            & \phantom{$<$}17\phantom{0}\\
 
3.2 $-$ 3.6 & 12           & 24            & \phantom{$<$}21\phantom{0}\\
 
3.6 $-$ 4.5 & 18           & 26            & \phantom{$<$}25\phantom{0}\\

\hline
\end{tabular}
\caption{The estimated inclusive, $\epem$ and $\rm h^+h^-$ backgrounds in each $\wgg$ bin.}
\label{bkgtable}
\end{center}
\end{table}

\begin{table}[H]
\begin{center}
\begin{tabular}{|c|c|c|c|c|} \hline 

$ W_{\gamma \gamma}$ & $\langle  W_{\gamma \gamma} \rangle$ & Number of& Background &$\sigma(\gamma \gamma \ra \ppbar)$ \\
(GeV) & (GeV) &Events & (\%) &(nb) \\ \hline
2.1 $-$ 2.2 & 2.15 &  216 &  \phantom{0}4 & 5.35 $\pm$ 0.36 $\pm$ 0.55\\

2.2 $-$ 2.3 & 2.25 &  252 &  \phantom{0}5 & 4.34 $\pm$ 0.27 $\pm$ 0.41\\

2.3 $-$ 2.4 & 2.35 &  182 &  \phantom{0}7 & 2.86 $\pm$ 0.21 $\pm$ 0.28\\

2.4 $-$ 2.5 & 2.45 &  111 &  10 & 1.78 $\pm$ 0.17 $\pm$ 0.18\\

2.5 $-$ 2.6 & 2.55 &   \phantom{0}61 & 16 & 1.01 $\pm$ 0.13 $\pm$ 0.11\\

2.6 $-$ 2.8 & 2.69 &   \phantom{0}57 & 25 & 0.50 $\pm$ 0.07 $\pm$ 0.09\\

2.8 $-$ 3.2 & 2.97 &   \phantom{0}32 & 54 & 0.12 $\pm$ 0.02 $\pm$ 0.05\\

3.2 $-$ 3.6 & 3.37 &   \phantom{0}15 & 57 & 0.06 $\pm$ 0.02 $\pm$ 0.03\\
 
3.6 $-$ 4.5 & 3.95 &   \phantom{0}12 & 69 & 0.02 $\pm$ 0.01 $\pm$ 0.01\\
\hline 
\end{tabular}

\caption{The number of events, estimated background and $\gamma \gamma
\ra \ppbar$ cross section as a function of $ W_{\gamma \gamma}$ for $
|\cos \theta^*| < 0.6$. The average value $\langle W_{\gamma \gamma}
\rangle$ of each bin corresponds to a weighted average according to a
$W_{\gamma \gamma}^{-12}$ distribution. The first uncertainty is
statistical, the second systematic. }
\label{res1}
\end{center}
\end{table}

\begin{table}[H]
\begin{center}

\begin{tabular}{|c | c | c |c|} \hline 
 & \multicolumn{3}{c|} {d$\sigma(\gamma \gamma \rightarrow \rm p \overline{p}) /  d|\cos\theta^*|$ (nb)  }  \\ \cline{2-4} 
 $|\cos\theta^*|$ & 2.1 GeV $ < W_{\gamma \gamma} < 2.5 $ GeV& 2.5 GeV $< W_{\gamma \gamma}   < 3.0 $ GeV & 3.0 GeV $< W_{\gamma \gamma} < 4.5 $ GeV \\ \hline
\phantom{0.}0 $-$ 0.1  & 10.36 $\pm$  0.64 $\pm$  1.08 & 0.86 $\pm$  0.16 $\pm$  0.14 & 0.06 $\pm$  0.02 $\pm$  0.03 \\
0.1 $-$ 0.2  &  \phantom{0}7.79 $\pm$  0.57 $\pm$  0.81 & 0.56 $\pm$  0.12 $\pm$  0.09 & 0.04 $\pm$  0.02 $\pm$  0.02 \\
0.2 $-$ 0.3  &  \phantom{0}6.15 $\pm$  0.53 $\pm$  0.64 & 0.62 $\pm$  0.14 $\pm$  0.10 & 0.03 $\pm$  0.02 $\pm$  0.02 \\
0.3 $-$ 0.4  &  \phantom{0}5.39 $\pm$  0.54 $\pm$  0.56 & 1.01 $\pm$  0.18 $\pm$  0.17 & 0.06 $\pm$  0.02 $\pm$  0.03 \\
0.4 $-$ 0.5  &  \phantom{0}3.77 $\pm$  0.50 $\pm$  0.39 & 0.82 $\pm$  0.18 $\pm$  0.14 & 0.07 $\pm$  0.03 $\pm$  0.04 \\
0.5 $-$ 0.6  &  \phantom{0}2.30 $\pm$  0.48 $\pm$  0.24 & 1.03 $\pm$  0.22 $\pm$  0.17 & 0.11 $\pm$  0.04 $\pm$  0.05 \\

\hline  
\end{tabular}
\caption{The differential cross section as a function of $|\cos\theta^*|$ for the different $\wgg$ ranges. The first uncertainty is statistical, the second systematic.  }
\label{difftable}
\end{center}
\end{table}

\begin{table}[H]
\begin{center}

\begin{tabular}{|c | c | c |c|} \hline 
& & \multicolumn{2}{c|} {$\sigma(\gamma \gamma \rightarrow \rm p \overline{p})$ (nb) }  \\  \cline{3-4}
 $\wgg$ (GeV) & $\langle \wgg \rangle $ (GeV) & $|\cos\theta^*| < 0.3$& $0.3 <  |\cos\theta^*| < 0.6$   \\ \hline
2.1 $-$ 2.2 & 2.15 &  (3.58 $\pm$  0.27 $\pm$  0.37) \phantom{$\times 10 ^{-3}$} &   
                      (1.78 $\pm$  0.27 $\pm$  0.18) \phantom{$\times 10 ^{-3}$}\\
2.2 $-$ 2.3 & 2.25 &  (2.95 $\pm$  0.21 $\pm$  0.28) \phantom{$\times 10 ^{-3}$}&   
                      (1.40 $\pm$  0.18 $\pm$  0.13) \phantom{$\times 10 ^{-3}$}\\
2.3 $-$ 2.4 & 2.35 &  (2.01 $\pm$  0.17 $\pm$  0.19) \phantom{$\times 10 ^{-3}$} & 
                      (0.85 $\pm$  0.13 $\pm$  0.08) \phantom{$\times 10 ^{-3}$}\\
2.4 $-$ 2.5 & 2.45  & (1.21 $\pm$  0.13 $\pm$  0.12) \phantom{$\times 10 ^{-3}$}& 
                      (0.57 $\pm$  0.11 $\pm$  0.06) \phantom{$\times 10 ^{-3}$}\\
2.5 $-$ 2.6 & 2.55  & (0.54 $\pm$  0.09 $\pm$  0.06) \phantom{$\times 10 ^{-3}$}& 
                      (0.47 $\pm$  0.09 $\pm$  0.05) \phantom{$\times 10 ^{-3}$}\\
2.6 $-$ 2.8 & 2.69  & (0.19 $\pm$  0.04 $\pm$  0.03) \phantom{$\times 10 ^{-3}$}& 
                      (0.31 $\pm$  0.06 $\pm$  0.05) \phantom{$\times 10 ^{-3}$}\\
2.8 $-$ 3.2 & 2.97  & (0.03 $\pm$  0.01 $\pm$  0.01) \phantom{$\times 10 ^{-3}$}&  
                      (0.09 $\pm$  0.02 $\pm$  0.04) \phantom{$\times 10 ^{-3}$}\\
3.2 $-$ 3.6 & 3.37  & (0.02 $\pm$  0.01 $\pm$  0.01) \phantom{$\times 10 ^{-3}$} &  
                      (0.04 $\pm$  0.01 $\pm$  0.02) \phantom{$\times 10 ^{-3}$}\\
3.6 $-$ 4.5 & 3.95  & (\phantom{0.}10 $\pm$  \phantom{0.0}4 $\pm$  \phantom{0.0}6) $\times 10 ^{-3}$ &  
                      (\phantom{0.}11 $\pm$  \phantom{0.0}5 $\pm$  \phantom{0.0}7) $\times 10 ^{-3}$ \\
\hline  
\end{tabular}
\caption{$\gamma \gamma \ra \ppbar$ cross section as a function of $ W_{\gamma \gamma}$ for $ |\cos \theta^*| < 0.3$ and $ 0.3 < |\cos \theta^*| < 0.6$. The average value $\langle W_{\gamma \gamma} \rangle$ of each bin corresponds to a weighted average according to a $W_{\gamma \gamma}^{-12}$ distribution. The first uncertainty is statistical, the second systematic.}
\label{diffwggtable}
\end{center}
\end{table}

%
%

\newpage

\newpage

\newpage

\newpage

\newpage

\newpage

\end{document}

%% file: namelist266.tex
\typeout{   }     
\typeout{Using author list for paper 261 -  }
\typeout{$Modified: Jul 15 2001 by smele $}
\typeout{!!!!  This should only be used with document option a4p!!!!}
\typeout{   }
%
%
%
%
%
%

\newcount\tutecount  \tutecount=0
\def\tutenum#1{\global\advance\tutecount by 1 \xdef#1{\the\tutecount}}
\def\tute#1{$^{#1}$}
\tutenum\aachen            
\tutenum\nikhef            
\tutenum\mich              
\tutenum\lapp              
\tutenum\basel             
\tutenum\lsu               
\tutenum\beijing           
\tutenum\bologna           
\tutenum\tata              
\tutenum\ne                
\tutenum\bucharest         
\tutenum\budapest          
\tutenum\mit               
\tutenum\panjab            
\tutenum\debrecen          
\tutenum\dublin            
\tutenum\florence          
\tutenum\cern              
\tutenum\wl                
\tutenum\geneva            
\tutenum\hefei             
\tutenum\lausanne          
\tutenum\lyon              
\tutenum\madrid            
\tutenum\florida           
\tutenum\milan             
\tutenum\moscow            
\tutenum\naples            
\tutenum\cyprus            
\tutenum\nymegen           
\tutenum\caltech           
\tutenum\perugia           
\tutenum\peters            
\tutenum\cmu               
\tutenum\potenza           
\tutenum\prince            
\tutenum\riverside         
\tutenum\rome              
\tutenum\salerno           
\tutenum\ucsd              
\tutenum\sofia             
\tutenum\korea             
\tutenum\purdue            
\tutenum\psinst            
\tutenum\zeuthen           
\tutenum\eth               
\tutenum\hamburg           
\tutenum\taiwan            
\tutenum\tsinghua          

{
\parskip=0pt
\noindent
{\bf The L3 Collaboration:}
\ifx\selectfont\undefined
 \baselineskip=10.8pt
 \baselineskip\baselinestretch\baselineskip
 \normalbaselineskip\baselineskip
 \ixpt
\else
 \fontsize{9}{10.8pt}\selectfont
\fi
\medskip
\tolerance=10000
\hbadness=5000
\raggedright
\hsize=162truemm\hoffset=0mm
\def\r{\rlap,}
\noindent

P.Achard\r\tute\geneva\ 
O.Adriani\r\tute{\florence}\ 
M.Aguilar-Benitez\r\tute\madrid\ 
J.Alcaraz\r\tute{\madrid}\ 
G.Alemanni\r\tute\lausanne\
J.Allaby\r\tute\cern\
A.Aloisio\r\tute\naples\ 
M.G.Alviggi\r\tute\naples\
H.Anderhub\r\tute\eth\ 
V.P.Andreev\r\tute{\lsu,\peters}\
% ...
\newpage
\begin{list}{A}{\itemsep=0pt plus 0pt minus 0pt\parsep=0pt plus 0pt minus 0pt
                \topsep=0pt plus 0pt minus 0pt}
\item[\aachen]
 III. Physikalisches Institut, RWTH, D-52056 Aachen, Germany$^{\S}$
\item[\nikhef] National Institute for High Energy Physics, NIKHEF, 
     and University of Amsterdam, NL-1009 DB Amsterdam, The Netherlands
\item[\mich] University of Michigan, Ann Arbor, MI 48109, USA
\item[\lapp] Laboratoire d'Annecy-le-Vieux de Physique des Particules, 
     LAPP,IN2P3-CNRS, BP 110, F-74941 Annecy-le-Vieux CEDEX, France
\item[\basel] Institute of Physics, University of Basel, CH-4056 Basel,
     Switzerland
\item[\lsu] Louisiana State University, Baton Rouge, LA 70803, USA
\item[\beijing] Institute of High Energy Physics, IHEP, 
  100039 Beijing, China$^{\triangle}$ 
\item[\bologna] University of Bologna and INFN-Sezione di Bologna, 
     I-40126 Bologna, Italy
\item[\tata] Tata Institute of Fundamental Research, Mumbai (Bombay) 400 005, India
\item[\ne] Northeastern University, Boston, MA 02115, USA
\item[\bucharest] Institute of Atomic Physics and University of Bucharest,
     R-76900 Bucharest, Romania
\item[\budapest] Central Research Institute for Physics of the 
     Hungarian Academy of Sciences, H-1525 Budapest 114, Hungary$^{\ddag}$
\item[\mit] Massachusetts Institute of Technology, Cambridge, MA 02139, USA
\item[\panjab] Panjab University, Chandigarh 160 014, India.
\item[\debrecen] KLTE-ATOMKI, H-4010 Debrecen, Hungary$^\P$
\item[\dublin] Department of Experimental Physics,
  University College Dublin, Belfield, Dublin 4, Ireland
\item[\florence] INFN Sezione di Firenze and University of Florence, 
     I-50125 Florence, Italy
\item[\cern] European Laboratory for Particle Physics, CERN, 
     CH-1211 Geneva 23, Switzerland
\item[\wl] World Laboratory, FBLJA  Project, CH-1211 Geneva 23, Switzerland
\item[\geneva] University of Geneva, CH-1211 Geneva 4, Switzerland
\item[\hefei] Chinese University of Science and Technology, USTC,
      Hefei, Anhui 230 029, China$^{\triangle}$
\item[\lausanne] University of Lausanne, CH-1015 Lausanne, Switzerland
\item[\lyon] Institut de Physique Nucl\'eaire de Lyon, 
     IN2P3-CNRS,Universit\'e Claude Bernard, 
     F-69622 Villeurbanne, France
\item[\madrid] Centro de Investigaciones Energ{\'e}ticas, 
     Medioambientales y Tecnol\'ogicas, CIEMAT, E-28040 Madrid,
     Spain${\flat}$ 
\item[\florida] Florida Institute of Technology, Melbourne, FL 32901, USA
\item[\milan] INFN-Sezione di Milano, I-20133 Milan, Italy
\item[\moscow] Institute of Theoretical and Experimental Physics, ITEP, 
     Moscow, Russia
\item[\naples] INFN-Sezione di Napoli and University of Naples, 
     I-80125 Naples, Italy
\item[\cyprus] Department of Physics, University of Cyprus,
     Nicosia, Cyprus
\item[\nymegen] University of Nijmegen and NIKHEF, 
     NL-6525 ED Nijmegen, The Netherlands
\item[\caltech] California Institute of Technology, Pasadena, CA 91125, USA
\item[\perugia] INFN-Sezione di Perugia and Universit\`a Degli 
     Studi di Perugia, I-06100 Perugia, Italy   
\item[\peters] Nuclear Physics Institute, St. Petersburg, Russia
\item[\cmu] Carnegie Mellon University, Pittsburgh, PA 15213, USA
\item[\potenza] INFN-Sezione di Napoli and University of Potenza, 
     I-85100 Potenza, Italy
\item[\prince] Princeton University, Princeton, NJ 08544, USA
\item[\riverside] University of Californa, Riverside, CA 92521, USA
\item[\rome] INFN-Sezione di Roma and University of Rome, ``La Sapienza",
     I-00185 Rome, Italy
\item[\salerno] University and INFN, Salerno, I-84100 Salerno, Italy
\item[\ucsd] University of California, San Diego, CA 92093, USA
\item[\sofia] Bulgarian Academy of Sciences, Central Lab.~of 
     Mechatronics and Instrumentation, BU-1113 Sofia, Bulgaria
\item[\korea]  The Center for High Energy Physics, 
     Kyungpook National University, 702-701 Taegu, Republic of Korea
\item[\purdue] Purdue University, West Lafayette, IN 47907, USA
\item[\psinst] Paul Scherrer Institut, PSI, CH-5232 Villigen, Switzerland
\item[\zeuthen] DESY, D-15738 Zeuthen, Germany
\item[\eth] Eidgen\"ossische Technische Hochschule, ETH Z\"urich,
     CH-8093 Z\"urich, Switzerland
\item[\hamburg] University of Hamburg, D-22761 Hamburg, Germany
\item[\taiwan] National Central University, Chung-Li, Taiwan, China
\item[\tsinghua] Department of Physics, National Tsing Hua University,
      Taiwan, China
\item[\S]  Supported by the German Bundesministerium 
        f\"ur Bildung, Wissenschaft, Forschung und Technologie
\item[\ddag] Supported by the Hungarian OTKA fund under contract
numbers T019181, F023259 and T037350.
\item[\P] Also supported by the Hungarian OTKA fund under contract
  number T026178.
\item[$\flat$] Supported also by the Comisi\'on Interministerial de Ciencia y 
        Tecnolog{\'\i}a.
\item[$\sharp$] Also supported by CONICET and Universidad Nacional de La Plata,
        CC 67, 1900 La Plata, Argentina.
\item[$\triangle$] Supported by the National Natural Science
  Foundation of China.
\end{list}
}
\vfill
